# Optimizing the 'For' loop: Comparison of 'For' loop and micro 'For' loop

Observing the orthogonal deficiency in 'for' loops to improve loop runtime.

Rishabh Jain
CETG-TIPBU
Cisco Systems India Pvt. Ltd.
Bangalore, India
rishajai@cisco.com

Sakshi Gupta
Security Business Group
Cisco Systems India Pvt. Ltd.
Bangalore, India
saksgupt@cisco.com

*Abstract*— **Looping is one of the fundamental logical instructions used for repeating a block of code. It is used in programs across all programming languages. Traditionally, in languages like C, the 'for' loop is used extensively for repeated execution of a block of code, due to its ease for use and simplified representation. This paper proposes a new way of representing the 'for' loop to improve its runtime efficiency and compares the experimental statistics with the traditional 'for' loop representation. It is found that for small number of iterations, the difference in computational time may not be considerable. But given any large number of iterations, the difference is noticeable.**

## I. Introduction

A loop is block of code which is intended to run multiple times. Loops have been part of programming since the beginning of structured code in assembly language. In the earliest form using GOTO statements in assembly code, it is has now evolved to much simpler logical abstractions like 'while' and 'for' loops.

Loops generally have three attributes to it, namely: variable initialization, running condition and variable update. The variable initialization, though optional, is the condition that the loop must satisfy before it can begin with the first iteration of the content of the loop. The variable update is optional.

The most important parameter for a looping condition is the running condition. It is this condition, when met with a 'true' does the loop continue with the next iteration. This condition is also optional, however, leaving this condition blank leads to the loop becoming an infinite loop. When the running condition statement returns a 'false', the next iteration is not begun and the loop statement is exited. This means that the code is allowed to proceed with the first statement occurring immediately after the loop statement. This paper makes a simple modification to the 'For' loop representation to improve the efficiency of the loop. The improvement is seen in the time used by the looping mechanism to work, regardless of the content.

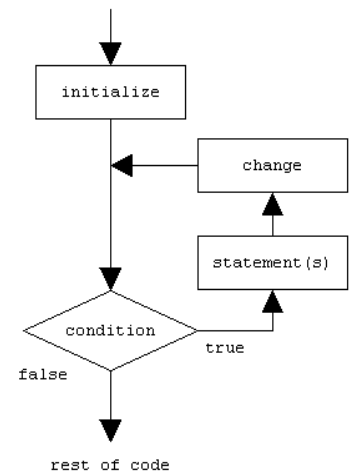

Fig. 1 Flow Chart of a traditional 'For' Loop. Note that the Fig. contains the three parts of the traditional loop elements: Variable Initialization, Running Condition, and Variable Update.

## II. Traditional Approach

The traditional approach of the 'for' loop, with context of C programming Language looks like this:

*for(InitializationExpression;TestExpression;Variable Update)*

In the traditional approach, typically, a variable is taken and initialized to a desired value which can be used both to check the Running conditions and update the variable. This would mean that the traditional application of 'for' loop looks like this:

*for( i=0; i<n; i++){*
*//Code logic*
*};*

## III. Proposed Approach

In the proposed approach, we incorporate the variable update in the running expression itself. Hence the usage becomes:

*for( i=0; i++<n;){*
*//Code Logic*
*};*

In this approach we attempt to syntactically eliminate the need of a variable update. Though, the logical usage remains the same. The difference in the above two representations is that if the variable 'i' is used in the code logic then the values at different times would be different. This is because in the traditional usage the variable 'i' gets updated after the iteration is complete, meanwhile, in the proposed representation the variable is updated before the iteration begins.

From now on we will be referring to the proposed approach as 'Micro For loop'.

### IV. COMPARISON OF WORKING

For comparing with the traditional approach, we check the assembly code equivalent of the two formats of representation.

```
.LFB0:
        movl    $0,         -8(%rbp)
        jmp     .L2
.L3:
        addl    $1,         -8(%rbp)
.L2:
        movl    -8(%rbp),   %eax
        cmpl    -4(%rbp),   %eax
        jl      .L3
```

Assembly code equivalent of the traditional approach. In this example the number of iterations are considered to be at -4(%rbp) and initial value to be at -8(%rbp). Compiler used is GNU GCC C.

```
.LFB0:
        movl    $0,         -8(%rbp)
        nop
.L2:
        movl    -8(%rbp),   %eax
        leal    1(%rax),    %edx
        movl    %edx,       -8(%rbp)
        cmpl    -4(%rbp),   %eax
        jl      .L2
```

Assembly code equivalent of the proposed approach. In this example the number of iterations are considered to be at -4(%rbp) and initial value to be at -8(%rbp). Compiler used is GNU GCC C.

The assembly code equivalent for the traditional approach shows the use of 2 jump statements, one of which being a conditional jump. In the traditional approach the number of clock cycles used per iteration of the 'for' loop will be higher since the condition is first checked, the loop body is executed and then the iterator is incremented. In contrast, the proposed approach shows the use of only one jump statement. The condition will be checked and the iterator is incremented before the loop body is executed saving clock cycles per iteration of the 'for' loop hence giving better performance.

The two approaches were compared theoretically by calculating the number of clock cycles utilized for the number of values of 'n', i.e. the number of iterations.

The jmp instruction can take from 23-32 clock cycles. The average was considered to do the calculations. The conditional jump, j jumps to an address set in a flag, it takes 25-33 cycles.

| Value of 'n' ($10^8$) | Micro loop (sec) | Traditional Loop (sec) | Difference (sec) |
|---|---|---|---|
| 0.0006 | 0.001333333 | 0.002233333 | 0.0009 |
| 0.0007 | 0.001555556 | 0.002605556 | 0.00105 |
| 0.0008 | 0.001777778 | 0.002977778 | 0.0012 |
| 0.01 | 0.022222222 | 0.037222222 | 0.015 |
| 0.03 | 0.066666667 | 0.111666667 | 0.045 |
| 0.05 | 0.111111111 | 0.186111111 | 0.075 |
| 1 | 2.222222222 | 3.722222222 | 1.5 |

It may be observed that theoretically the efficiency can be expected to be up by 40.2%, although because of variations in runtime environment and process execution queue the results might vary experimentally.

### V. EXPERIMENTAL RESULTS

It is found out when experimentally running the two 'for' loops in similar running environments that it indeed gives better results when micro for loop is run over the traditional 'for' loop.

| Value of n ($10^8$) | For loop (ms) | Micro For Loop (ms) | Efficiency (%) |
|---|---|---|---|
| 0.5 | 140 | 120 | 14.28 |
| 1 | 290 | 250 | 13.79 |
| 2 | 590 | 500 | 15.25 |
| 3 | 880 | 750 | 14.77 |
| 4 | 1170 | 1000 | 14.53 |
| 5 | 1460 | 1270 | 13.01 |
| 6 | 1750 | 1530 | 12.57 |
| 7 | 2040 | 1780 | 12.74 |
| 8 | 2370 | 2029 | 14.38 |
| 9 | 2660 | 2300 | 13.53 |
| 10 | 2960 | 2310 | 21.96 |
| 20 | 5330 | 4757 | 10.75 |
| 30 | 5627 | 5291 | 5.97 |
| 40 | 5537 | 5271 | 4.80 |

Using these results, we find the statistical results for the increase in efficiency,

*Standard Deviation = 4.08*
*Variance = 16.66*
*Populated Standard Deviation = 3.93*

*Populated Variance = 15.47*

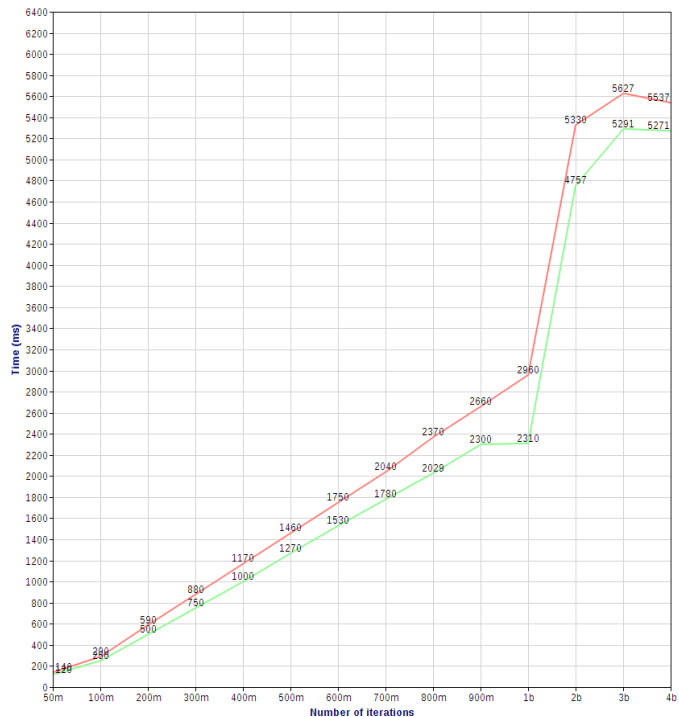

It can be seen that there is on an average increase in performance by ~ 13%. This can be considerable for loops that run in several million iterations, such as ones employed by real-time systems, video codecs, etc.

## Conclusion

The paper proposed a new way of representing the 'for' loop to improve its runtime efficiency. The experimental and theoretical statistics for the traditional 'for' loop representation and the proposed 'for' loop representation was compared. It is found that for small number of iterations, the difference in computational time may not be considerable. But given any large number of iterations, the difference is noticeable. It was observed that theoretically the efficiency can be 40.2%, although because of variations in runtime environment and process execution queue the results might vary experimentally. Experimentally, an average increase in performance of 13% was observed. This increase in considerable for loops run in several million iterations.

## Acknowledgment

It would be befitting to thank Mr. Rajat Tandon for his observations and Ms. Rachna Gupta for her contribution in collecting valuable data used in this paper. We also thank Anupam Sarda, Sai Prasad Nooka and Ranjini Subramanium for the motivation.